\documentclass[aps,prl,reprint,nolongbibliography,noeprint,superscriptaddress]{revtex4-2}
\usepackage{amsfonts,amssymb,amsmath,bm,graphicx,color}
\usepackage[rightcaption]{sidecap}
\usepackage[colorlinks=true,citecolor=blue,linkcolor=blue,urlcolor=blue]{hyperref}
\allowdisplaybreaks[4]
\sidecaptionvpos{figure}{c}
\DeclareMathOperator{\im}{Im}
\DeclareMathOperator{\tr}{tr}
\newcommand{\hb}[1]{\hat{\bm{#1}}}
\newcommand{\hc}[1]{\hat{\mathcal{#1}}}

\begin{document}
%--- Front Matter
\title{Spin accumulation without spin current}
\author{Atsuo Shitade}
\affiliation{Institute for Molecular Science, Aichi 444-8585, Japan}
\author{Gen Tatara}
\affiliation{RIKEN Center for Emergent Matter Science (CEMS) and RIKEN Cluster for Pioneering Research (CPR), 2-1 Hirosawa, Wako, Saitama 351-0198, Japan}
\date{\today}
\begin{abstract}
  The spin Hall (SH) effect is a phenomenon in which the spin current flows perpendicular to an applied electric field and causes the spin accumulation at the boundaries.
  However, in the presence of spin-orbit couplings, the spin current is not well defined.
  Here, we calculate the spin response to an electric-field gradient, which naturally appears at the boundaries.
  We derive a generic formula using the Bloch wave functions and the phenomenological relaxation time.
  We also calculate the response for the uniform Rashba model with $\delta$-function nonmagnetic disorder within the first-order Born approximation and corresponding vertex corrections.
  We find the nonzero spin accumulation, although the SH conductivity exactly vanishes.
\end{abstract}
\maketitle
%--- Main Matter
\paragraph{Introduction.}
Spintronics is an active research field in condensed-matter physics to make use of the spin degree of freedom of electrons.
Key steps are creation, transportation, and detection of spins, and, hence, the spin current has been believed to play an important role.
Such a current can be generated  via spin-orbit (SO) couplings perpendicular to an applied electric field.
This phenomenon, proposed by D'yakonov and Perel'~\cite{Dyakonov1971} and later by Hirsch~\cite{PhysRevLett.83.1834}, is called the spin Hall (SH) effect~\cite{RevModPhys.87.1213}.
It has attracted renewed interest since the theoretical proposals~\cite{Murakami1348,PhysRevLett.92.126603} and experimental observations in semiconductors~\cite{PhysRevLett.94.047204,Kato1910}.

\begin{SCfigure*}
  \centering
  \includegraphics[clip,width=0.55\textwidth]{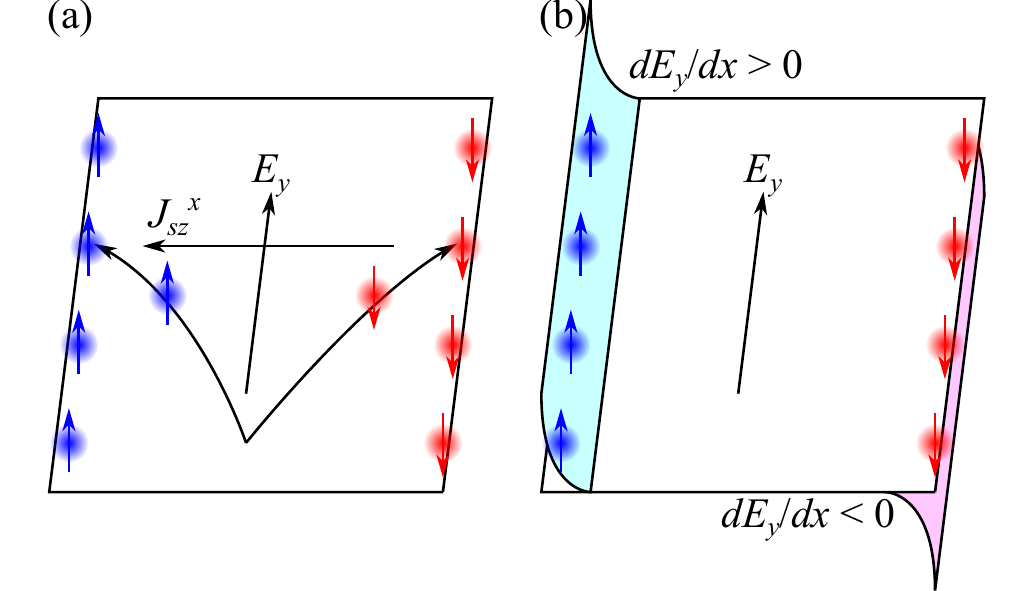}
  \caption{%
  (a) Typical scenario:
  The spin current is generated by a uniform electric field via the SH effect and then turns into spin at the boundaries.
  In the uniform Rashba model, the SH conductivity vanishes, and no spin accumulation is expected.
  (b) Our scenario:
  Spin is induced by the electric-field gradient at the boundaries.
  The spin accumulation may occur even when the SH conductivity vanishes.
  Our theory is free from the ambiguity regarding the definition of the spin current and spin torque.
  } \label{fig:she}
\end{SCfigure*}

Experimentally, the spin current in the SH effect has not been directly observed.
Only indirectly observed are the charge current in the inverse SH effect~\cite{1.2199473,nature04937} and the magnetization dynamics in the ferromagnetic resonance~\cite{PhysRevLett.101.036601,PhysRevLett.106.036601}.
In Refs.~\cite{PhysRevLett.94.047204,Kato1910}, the spin accumulation at the boundaries was detected optically and attributed to the SH effect:
The spin current is generated via the SH effect and then turns into spin at the boundaries, as depicted in Fig.~\ref{fig:she}(a).
Hence, spin, rather than the spin current, is the primary physical object in order to describe these experimental results.
This idea was pointed out already in the first theoretical proposal~\cite{Dyakonov1971}
and repeatedly in many subsequent papers~\cite{PhysRevLett.93.226602,PhysRevB.70.195343,PhysRevLett.95.046601,PhysRevB.72.081301,PhysRevB.73.075303,RASHBA200631,PhysRevB.74.035340,PhysRevLett.99.106601,PhysRevLett.102.196802,PhysRevB.81.113304,PhysRevLett.109.246604}.

In the presence of SO couplings, spin is not conserved, and the spin current is not well defined.
When a spin current density $J_{sa}^{\phantom{sa} i}(t, \bm{x})$ is given, there exists the corresponding spin torque density $\tau_{a}(t, \bm{x})$, and the spin continuity equation is expressed as
$\partial_{t} s_{a}(t, \bm{x}) + \partial_{x^{i}} J_{sa}^{\phantom{sa} i}(t, \bm{x})  = \tau_{a}(t, \bm{x})$.
Widely used is the conventional definition, $\hat{J}_{sa}^{\phantom{sa} i}(\bm{k}) = \{\hat{s}_{a}, \hat{v}^{i}(\bm{k})\}/2$,
where $\hat{s}_{a}$ and $\hat{v}^{i}(\bm{k})$ are the spin and velocity operators, respectively.
However, this definition is unphysical in the sense that its uniform equilibrium expectation value is nonzero in noncentrosymmetric systems, such as the Rashba and Dresselhaus models~\cite{PhysRevB.68.241315}.
Another definition is the so-called conserved spin current~\cite{PhysRevLett.96.076604,PhysRevB.77.075304}.
If the spin torque vanishes in average over the whole system, we can define the spin torque dipole density as $\tau_{a}(t, \bm{x}) = -\partial_{x^{i}} P_{\tau a}^{\phantom{\tau a} i}(t, \bm{x})$,
and $\tilde{J}_{sa}^{\phantom{sa} i}(t, \bm{x}) = J_{sa}^{\phantom{sa} i}(t, \bm{x}) + P_{\tau a}^{\phantom{\tau a} i}(t, \bm{x})$ is conserved on average.
This definition has interesting properties, such as the St\v{r}eda formula between the SH conductivity and the SO magnetic susceptibility~\cite{PhysRevLett.97.236805}
and the Mott relation between the SH and the spin Nernst conductivities~\cite{PhysRevB.98.081401,PhysRevB.104.L241411}.
Whatever definition we choose, however, we need to consider the corresponding spin torque density to evaluate the observable spin density. 
Note that using the scattering approach for mesoscopic systems,
the prohibition of the equilibrium spin current~\cite{PhysRevB.71.153315}, an electrical measurement scheme~\cite{PhysRevLett.106.206602}, and the Onsager reciprocal relations~\cite{PhysRevB.86.155118}
were shown without defining the spin current.

In the case of the Rashba model that describes $n$-type semiconductor heterostructures, the SH conductivity of the conventional spin current exactly vanishes
when the vertex corrections are taken into account~\cite{PhysRevB.70.041303,PhysRevLett.93.226602,PhysRevB.71.033311,PhysRevB.71.245318,PhysRevB.73.049901,PhysRevB.71.245327,PhysRevLett.96.056602}.
This cancellation is owing to the special property
that the conventional spin current operator is proportional to the time derivative of the spin operator~\cite{PhysRevB.71.245318,PhysRevB.73.049901,PhysRevB.71.245327}.
The SH conductivity of the conserved spin current also vanishes~\cite{PhysRevB.73.113305}.
Following the typical scenario in Fig.~\ref{fig:she}(a), the spin accumulation would be zero but, in fact, observed experimentally~\cite{Kato1910}.
Thus, it is clearly insufficient to focus on the SH conductivity only.

In contrast to the spin current, spin is well defined.
Regarding the Rashba model, the spin polarization at the boundaries has been calculated
using the coupled diffusion equations obtained microscopically~\cite{PhysRevLett.93.226602,PhysRevB.70.195343,RASHBA200631,PhysRevB.74.035340},
the Landauer-Keldysh formalism~\cite{PhysRevLett.95.046601,PhysRevB.72.081301,PhysRevB.73.075303},
and the scattering problem~\cite{PhysRevLett.99.106601,PhysRevLett.102.196802,PhysRevB.81.113304}.
Now it is well understood that the essence of the SH effect is the spin accumulation.
However, in these formalisms, we need to impose the open boundary conditions or attach the leads to the system.
It is difficult to deal with such finite geometries in first-principles calculations for real materials, which may have multiple bands and complicated SO couplings.
Hence, the Kubo formula of the SH conductivity is widely used in first-principles calculations~\cite{PhysRevLett.94.226601,PhysRevLett.95.156601,PhysRevLett.106.056601} despite the aforementioned problems.
It is highly desired to establish the Kubo formula of the spin accumulation.

Recently, one of the authors considered the spin response to an electric-field gradient~\cite{PhysRevB.98.174422}.
When a uniform electric field is applied to a finite-size system, the charge current vanishes at the boundaries.
What we call the electric field here effectively describes such a boundary effect, and its gradient has peaks there as depicted in Fig.~\ref{fig:she}(b).
Since the spin-diffusion length that characterizes the spin accumulation is much longer than the mean free path,
we can safely assume that the electric field slowly decreases towards the boundaries.
Then, the spin accumulation can be emulated imposing the periodic boundary conditions, which are compatible to first-principles calculations.
The theory also explains generation of spin current using the SH effect or the spin pumping and detection using the inverse SH effect in terms of the nonlocal spin fluctuation.

In this Letter, we study the spin response to the electric-field gradient with the quantum-mechanical linear-response theory.
First, we derive a generic formula expressed by the Bloch wave functions.
Although disorder effects are taken into account via a phenomenological relaxation time, the formula can be applied to any Bloch Hamiltonian.
Second, we calculate the spin response with the Green's functions.
We consider the uniform Rashba model with $\delta$-function nonmagnetic disorder within the first Born approximation and corresponding vertex corrections,
which results in the vanishing SH conductivity~\cite{PhysRevB.70.041303,PhysRevLett.93.226602,PhysRevB.71.033311,PhysRevB.71.245318,PhysRevB.73.049901,PhysRevB.71.245327,PhysRevLett.96.056602}.
Nonetheless, we find the nonzero spin accumulation, which is consistent with the experimental result~\cite{Kato1910}.
This theory enables us to calculate the observable quantity in the SH effect for real materials.

\paragraph{Bloch formulas.}
First, we calculate the spin--charge-current correlation function that characterizes
$\langle \Delta \hat{s}_{a} \rangle(\Omega, \bm{Q}) = \chi_{\hat{s}_{a} \hat{J}^{j}}^{\mathrm{R}}(\Omega, \bm{Q}) A_{j}(\Omega, \bm{Q})$,
where $\Omega$ and $\bm{Q}$ are the frequency and the wave number of an external vector potential.
Using the Bloch wave-functions $| u_{n}(\bm{k}) \rangle$ for the Bloch Hamiltonian $\hc{H}(\bm{k})$, the correlation function is expressed as
\begin{align}
  \chi_{\hat{s}_{a} \hat{J}^{j}}^{\mathrm{R}}(\Omega, \bm{Q})
  = & -q \sum_{nm} \int \frac{d^{d} k}{(2 \pi)^{d}}
  \langle u_{n}(\bm{k}_{-}) | \hat{s}_{a} | u_{m}(\bm{k}_{+}) \rangle \notag \\
  & \times \langle u_{m}(\bm{k}_{+}) | \hat{v}^{j}(\bm{k}; \bm{Q}) | u_{n}(\bm{k}_{-}) \rangle \notag \\
  & \times \frac{f(\epsilon_{n}(\bm{k}_{-})) - f(\epsilon_{m}(\bm{k}_{+}))}{\hbar \Omega + \epsilon_{n}(\bm{k}_{-}) - \epsilon_{m}(\bm{k}_{+}) + i \eta} \notag \\
  = & \chi_{\hat{s}_{a} \hat{J}^{j}}(0, \bm{Q}) + (i \Omega) \alpha_{\hat{s}_{a} \hat{J}^{j}}^{\mathrm{R}}(\Omega, \bm{Q}), \label{eq:so_bloch1}
\end{align}
where $q$ is the electron charge, $d$ is the spatial dimension, $\eta \rightarrow +0$ is the convergence factor,
$\hat{v}^{j}(\bm{k}; \bm{Q}) = [\hat{v}^{j}(\bm{k}_{+}) + \hat{v}^{j}(\bm{k}_{-})]/2$ with $\bm{k}_{\pm} = \bm{k} \pm \bm{Q}/2$,
and $f(\epsilon) = [e^{(\epsilon - \mu)/T} + 1]^{-1}$ is the Fermi distribution function.
We expand Eq.~\eqref{eq:so_bloch1} up to the first order with respect to $\bm{Q}$ with keeping $\Omega$ nonzero.
The first term in Eq.~\eqref{eq:so_bloch1} takes the form of $\chi_{\hat{s}_{a} \hat{J}^{j}}(0, \bm{Q}) = \epsilon^{ijk} (i Q_{i}) \chi_{ak}^{\mathrm{so}}$,
and we reproduce the SO magnetic susceptibility~\cite{PhysRevResearch.3.023098,suppl},
\begin{align}
  \chi_{ak}^{\mathrm{so}}
  = & -\frac{q}{\hbar}
  \sum_{n} \int \frac{d^{d} k}{(2 \pi)^{d}}
  [(-\epsilon_{ijk} s_{na}^{\phantom{na} i} \partial_{k_{j}} \epsilon_{n} + s_{na} m_{nk}) f^{\prime}(\epsilon_{n}) \notag \\
  & + b_{nak} f(\epsilon_{n})]. \label{eq:so_bloch4}
\end{align}
The argument of $\bm{k}$ is omitted for simplicity.
We have introduced $s_{na} = \langle u_{n} | \hat{s}_{a} | u_{n} \rangle$, the magnetic moment $m_{nk}$~\cite{PhysRevB.59.14915,PhysRevLett.95.137204,PhysRevLett.95.169903},
spin magnetic quadrupole moment $s_{na}^{\phantom{na} i}$~\cite{PhysRevB.97.134423,PhysRevB.99.024404},
and spin Berry curvature $b_{nak}$ as
\begin{subequations} \begin{align}
  \epsilon^{ijk} m_{nk}
  = & \im [\langle \partial_{k_{i}} u_{n} | (\epsilon_{n} - \hc{H}) | \partial_{k_{j}} u_{n} \rangle], \label{eq:prop_bloch5a} \\
%%%%%%%%%%
  s_{na}^{\phantom{na} i}
  = & \im (\langle \partial_{k_{i}} u_{n} | \hat{Q}_{n} \hat{s}_{a} | u_{n} \rangle), \label{eq:prop_bloch5b} \\
%%%%%%%%%%
  \epsilon^{ijk} b_{nak}
  = & -\im [\langle \partial_{k_{i}} u_{n} | \hat{Q}_{n} (s_{na} + \hat{s}_{a}) \hat{Q}_{n} | \partial_{k_{j}} u_{n} \rangle]
  + \sum_{m (\not= n)} \notag \\
  & \times \frac{\im [\langle u_{n} | \hat{s}_{a} | u_{m} \rangle \langle u_{m} | (\partial_{k_{i}} \epsilon_{n} + \hbar \hat{v}^{i}) \hat{Q}_{n} | \partial_{k_{j}} u_{n} \rangle]}{\epsilon_{n} - \epsilon_{m}} \notag \\
  & - (i \leftrightarrow j), \label{eq:prop_bloch5c}
\end{align} \label{eq:prop_bloch5}\end{subequations}
with $\hat{Q}_{n} = 1 - | u_{n} \rangle \langle u_{n} |$ being the antiprojection operator.
Equation~\eqref{eq:prop_bloch5c} is a spin analog of the Berry curvature
because it is reduced to the Berry curvature when $\hat{s}_{a}$ is replaced by $1$
and totally antisymmetric with respect to $\partial_{k_{i}}$, $\partial_{k_{j}}$, and $\partial_{B^{a}}$.
Here, $B^{a}$ is the Zeeman field conjugate to $\hat{s}_{a}$.

The second term in Eq.~\eqref{eq:so_bloch1} takes the form of~\cite{suppl}
\begin{align}
  \alpha_{\hat{s}_{a} \hat{J}^{j}}^{\mathrm{R}}(\Omega, \bm{Q})
  = & [\chi_{\hat{s}_{a} \hat{J}^{j}}^{\mathrm{R}}(\Omega, \bm{Q}) - \chi_{\hat{s}_{a} \hat{J}^{j}}(0, \bm{Q})]/(i \Omega) \notag \\
  = & \frac{i \hbar}{\hbar \Omega + i \eta}  \alpha_{a}^{\phantom{a} j}
  + \frac{\hbar^{2} (i Q_{i})}{(\hbar \Omega + i \eta)^{2}} \gamma_{a}^{\phantom{a} ij (\mathrm{I})} \notag \\
  & - \frac{i \hbar (i Q_{i})}{\hbar \Omega + i \eta} \gamma_{a}^{\phantom{a} ij (\mathrm{II})}, \label{eq:spin_bloch2}
\end{align}
where
\begin{subequations} \begin{align}
  \alpha_{a}^{\phantom{a} j}
  = & -\frac{q}{\hbar}
  \sum_{n} \int \frac{d^{d} k}{(2 \pi)^{d}}
  s_{na} \partial_{k_{j}} \epsilon_{n} f^{\prime}(\epsilon_{n}), \label{eq:spin_bloch3b} \\
%%%%%%%%%%
  \gamma_{a}^{\phantom{a} ij (\mathrm{I})}
  = & -\frac{q}{\hbar^{2}}
  \sum_{n} \int \frac{d^{d} k}{(2 \pi)^{d}}
  s_{na} \partial_{k_{i}} \epsilon_{n} \partial_{k_{j}} \epsilon_{n} f^{\prime}(\epsilon_{n}), \label{eq:spin_bloch3c} \\
%%%%%%%%%%
  \gamma_{a}^{\phantom{a} ij (\mathrm{II})}
  = & -\frac{q}{\hbar}
  \sum_{n} \int \frac{d^{d} k}{(2 \pi)^{d}} \notag \\
  & \times (s_{na}^{\phantom{na} i} \partial_{k_{j}} \epsilon_{n} - s_{na} \epsilon^{ijk} m_{nk}) f^{\prime}(\epsilon_{n}). \label{eq:spin_bloch3d}
\end{align} \label{eq:spin_bloch3}\end{subequations}
Equation~\eqref{eq:spin_bloch3b} describes the Edelstein effect~\cite{Edelstein1990233},
whereas Eqs.~\eqref{eq:spin_bloch3c} and \eqref{eq:spin_bloch3d} describe the spin accumulation induced by the electric-field gradient.
Note that we drop the interband Fermi-sea term because it breaks the time-reversal symmetry.

Combining Eqs.~\eqref{eq:so_bloch4} and \eqref{eq:spin_bloch2}, the spin density is induced by electromagnetic fields as
\begin{align}
  \langle \Delta \hat{s}_{a} \rangle(\Omega, \bm{Q})
  = & \frac{i \hbar}{\hbar \Omega + i \eta} \alpha_{a}^{\phantom{a} j} E_{j}(\Omega, \bm{Q}) \notag \\
  & + \left[\frac{\hbar^{2}}{(\hbar \Omega + i \eta)^{2}} \gamma_{a}^{\phantom{a} ij (\mathrm{I})} - \frac{i \hbar}{\hbar \Omega + i \eta} \gamma_{a}^{\phantom{a} ij (\mathrm{II})}\right] \notag \\
  & \times (i Q_{i}) E_{j}(\Omega, \bm{Q})
  + \chi_{ak}^{\mathrm{so}} B^{k}(\Omega, \bm{Q}). \label{eq:spin_bloch3a}
\end{align}
Taking the limit of $\Omega \rightarrow 0$ and introducing the phenomenological relaxation time $\hbar/\eta$, we arrive at one of our main results,
\begin{align}
  \langle \Delta \hat{s}_{a} \rangle(0, \bm{Q})
  = & \frac{\hbar}{\eta} \alpha_{a}^{\phantom{a} j} E_{j}(0, \bm{Q})
  - \left[\frac{\hbar^{2}}{\eta^{2}} \gamma_{a}^{\phantom{a} ij (\mathrm{I})} + \frac{\hbar}{\eta} \gamma_{a}^{\phantom{a} ij (\mathrm{II})}\right] \notag \\
  & \times (i Q_{i}) E_{j}(0, \bm{Q}). \label{eq:spin_bloch4}
\end{align}

Let us apply the above formulas to the uniform Rashba model,
\begin{equation}
  \hc{H}(\bm{k})
  = \frac{\hbar^{2} k^{2}}{2 m} + \hbar \alpha (k_{y} \sigma_{x} - k_{x} \sigma_{y}), \label{eq:rashba1}
\end{equation}
where $\bm{\sigma}$ is the Pauli matrix corresponding to the spin operator $\hb{s} = (\hbar/2) \bm{\sigma}$.
The eigenvalues are $\epsilon_{\sigma}(\bm{k}) = \hbar^{2} k^{2}/2 m + \sigma \hbar \alpha k$.
At $T = 0$, we obtain the SO magnetic susceptibility~\eqref{eq:so_bloch4} and spin accumulation~\eqref{eq:spin_bloch3d} as~\cite{suppl}
\begin{subequations} \begin{align}
  \chi_{zz}^{\mathrm{so}}
  = & -\frac{q}{4 \pi}
  \begin{cases}
    0 & (\mu > 0) \\
    \sqrt{1 + 2 \mu/m \alpha^{2}} & (\mu < 0)
  \end{cases}, \label{eq:rashba3c} \\
%%%%%%%%%%
  \gamma_{z}^{\phantom{z} xy (\mathrm{II})}
  = & -\frac{q}{8 \pi}
  \begin{cases}
    1 & (\mu > 0) \\
    0 & (\mu < 0)
  \end{cases}. \label{eq:rashba3d}
\end{align} \label{eq:rashba3}\end{subequations}
Note that Eq.~\eqref{eq:rashba3c} is consistent with the previous result~\cite{Bychkov1984,PhysRevB.94.085303}.
The spin accumulation~\eqref{eq:rashba3d} is nonzero when the chemical potential is above the Rashba crossing.
However, it is natural to ask if Eq.~\eqref{eq:rashba3d} survives when the vertex corrections are taken into account.

\paragraph{Green's functions.}
The above results are phenomenological in the sense that $\hbar/\eta$ is interpreted as the relaxation time.
Here, we consider $\delta$-function nonmagnetic disorder within the first Born approximation
and take into account the corresponding ladder-type vertex corrections for the uniform Rashba model~\eqref{eq:rashba1}.
The bare retarded Green's function is expressed as
\begin{align}
  \hat{g}^{\mathrm{R}}(\epsilon, \bm{k})
  = & \frac{1}{\epsilon + i \eta - \hc{H}(\bm{k})} \notag \\
  = & \frac{1}{2} [g_{+}^{\mathrm{R}}(\epsilon, \bm{k}) + g_{-}^{\mathrm{R}}(\epsilon, \bm{k})]
  + \frac{1}{2} [g_{+}^{\mathrm{R}}(\epsilon, \bm{k}) - g_{-}^{\mathrm{R}}(\epsilon, \bm{k})] \notag \\
  & \times (\sigma_{x} \sin \phi - \sigma_{y} \cos \phi), \label{eq:green1}
\end{align}
with $g_{\sigma}^{\mathrm{R}}(\epsilon, \bm{k}) = [\epsilon + i \eta - \epsilon_{\sigma}(\bm{k})]^{-1}$ being the diagonalized one.
The imaginary part of the self-energy is then expressed as
\begin{align}
  \hat{\Gamma}(\epsilon)
  = & -\im \left[n_{\mathrm{i}} v_{\mathrm{i}}^{2} \int \frac{d^{2} k}{(2 \pi)^{2}} \hat{g}^{\mathrm{R}}(\epsilon, \bm{k})\right] \notag \\
  = & \Gamma_{0}
  \begin{cases}
    1 & (\epsilon > 0) \\
    1/\sqrt{1 + 2 \epsilon/m \alpha^{2}}& (\epsilon < 0)
  \end{cases}, \label{eq:green2}
\end{align}
with $\Gamma_{0} = m n_{\mathrm{i}} v_{\mathrm{i}}^{2}/2 \hbar^{2}$.
Below, we denote $\hat{\Gamma}(\epsilon)$ as $\Gamma(\epsilon)$.
The renormalized retarded Green's function is expressed as
\begin{align}
  \hat{G}^{\mathrm{R}}(\epsilon, \bm{k})
  = & \frac{1}{\epsilon + i \Gamma(\epsilon) - \hc{H}(\bm{k})} \notag \\
  = & \frac{1}{2} [G_{+}^{\mathrm{R}}(\epsilon, \bm{k}) + G_{-}^{\mathrm{R}}(\epsilon, \bm{k})] \notag \\
  & + \frac{1}{2} [G_{+}^{\mathrm{R}}(\epsilon, \bm{k}) - G_{-}^{\mathrm{R}}(\epsilon, \bm{k})] \notag \\
  & \times (\sigma_{x} \sin \phi - \sigma_{y} \cos \phi), \label{eq:green4}
\end{align}
with $G_{\sigma}^{\mathrm{R}}(\epsilon, \bm{k}) = [\epsilon + i \Gamma(\epsilon) - \epsilon_{\sigma}(\bm{k})]^{-1}$~\cite{PhysRevB.94.035306,PhysRevB.95.245302}.

\begin{figure*}
  \centering
  \includegraphics[clip,width=0.98\textwidth]{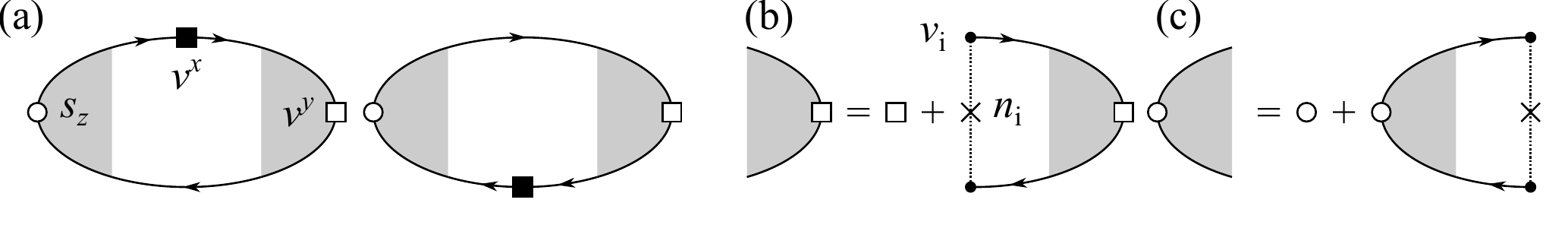}
  \caption{%
  Feynman diagrams for (a) the spin--charge-current correlation function of the first order with respect to $Q_{x}$,
  (b) the renormalized velocity vertex,
  and (c) the renormalized spin vertex.
  The filled squares, open squares, and open circles represent the bare vertices of $\hat{v}^{x}$, $\hat{v}^{y}$, and $\hat{s}_{z}$, respectively.
  } \label{fig:diagram}
\end{figure*}

\begin{widetext}
Now we evaluate the spin response to a vector potential,
\begin{align}
  \langle \Delta \hat{s}_{z} \rangle(\Omega, \bm{Q})
  = & i q A_{y}(\Omega, \bm{Q})
  \int \frac{d \epsilon}{2 \pi} \int \frac{d^{2} k}{(2 \pi)^{2}}
  \tr [\hat{s}_{z} \hat{G}(\epsilon_{+}, \bm{k}_{+}) \hat{v}^{y}(\bm{k}; \bm{Q}) \hat{G}(\epsilon_{-}, \bm{k}_{-})]^{<} \notag \\
  = & i q A_{y}(\Omega, \bm{Q})
  \int \frac{d \epsilon}{2 \pi} \int \frac{d^{2} k}{(2 \pi)^{2}}
  \tr \{-\hat{s}_{z} \hat{G}^{\mathrm{R}}(\epsilon_{+}, \bm{k}_{+}) \hat{v}^{y}(\bm{k}; \bm{Q}) \hat{G}^{\mathrm{A}}(\epsilon_{-}, \bm{k}_{-}) [f(\epsilon_{+}) - f(\epsilon_{-})] \notag \\
  & + \hat{s}_{z} \hat{G}^{\mathrm{A}}(\epsilon_{+}, \bm{k}_{+}) \hat{v}^{y}(\bm{k}; \bm{Q}) \hat{G}^{\mathrm{A}}(\epsilon_{-}, \bm{k}_{-}) f(\epsilon_{+})
  - \hat{s}_{z} \hat{G}^{\mathrm{R}}(\epsilon_{+}, \bm{k}_{+}) \hat{v}^{y}(\bm{k}; \bm{Q}) \hat{G}^{\mathrm{R}}(\epsilon_{-}, \bm{k}_{-}) f(\epsilon_{-})\}, \label{eq:conv_green1}
\end{align}
in which $\epsilon_{\pm} = \epsilon \pm \hbar \Omega/2$, up to the first order with respect to $\Omega$ and $Q_{x}$.
The zeroth-order terms with respect to $Q_{x}$ vanish owing to the $C_{4}$ symmetry of the Rashba model.
The first-order terms are decomposed into two;
one is the zeroth-order Fermi-sea term with respect to $\Omega$ and describes the SO magnetic susceptibility,
whereas the other is the first-order Fermi-surface term and describes the spin accumulation.
These terms are expressed as~\cite{suppl}
\begin{subequations} \begin{align}
  \langle \Delta \hat{s}_{z} \rangle^{(0, 1, \mathrm{II})}(\Omega, \bm{Q})
  = & \frac{i \hbar q}{2} Q_{x} A_{y}(\Omega, \bm{Q})
  \int \frac{d \epsilon}{2 \pi} f(\epsilon) \int \frac{d^{2} k}{(2 \pi)^{2}} \notag \\
  & \times \tr [\hat{s}_{z} \hat{G}^{\mathrm{A}} (\hat{v}^{x} \hat{G}^{\mathrm{A}} \hat{v}^{y} - \hat{v}^{y} \hat{G}^{\mathrm{A}} \hat{v}^{x}) \hat{G}^{\mathrm{A}}
  - (\mathrm{A} \rightarrow \mathrm{R})], \label{eq:spin_green2a} \\
%%%%%%%%%%
  \langle \Delta \hat{s}_{z} \rangle^{(1, 1, \mathrm{I})}(\Omega, \bm{Q})
  = & \frac{i \hbar^{2} q}{4} \Omega Q_{x} A_{y}(\Omega, \bm{Q})
  \int \frac{d \epsilon}{2 \pi} f^{\prime}(\epsilon) \int \frac{d^{2} k}{(2 \pi)^{2}} \notag \\
  & \times \tr [-2 \hat{S}_{z} \hat{G}^{\mathrm{R}} (\hat{v}^{x} \hat{G}^{\mathrm{R}} \hat{V}^{y} - \hat{V}^{y} \hat{G}^{\mathrm{A}} \hat{v}^{x}) \hat{G}^{\mathrm{A}} \notag \\
  & + \hat{s}_{z} \hat{G}^{\mathrm{A}} (\hat{v}^{x} \hat{G}^{\mathrm{A}} \hat{v}^{y} - \hat{v}^{y} \hat{G}^{\mathrm{A}} \hat{v}^{x}) \hat{G}^{\mathrm{A}}
  + \hat{s}_{z} \hat{G}^{\mathrm{R}} (\hat{v}^{x} \hat{G}^{\mathrm{R}} \hat{v}^{y} - \hat{v}^{y} \hat{G}^{\mathrm{R}} \hat{v}^{x}) \hat{G}^{\mathrm{R}}], \label{eq:spin_green2b}
\end{align} \label{eq:spin_green2}\end{subequations}
\end{widetext}
and diagrammatically represented in Fig.~\ref{fig:diagram}(a).
The arguments of $\epsilon$ and $\bm{k}$ are omitted for simplicity.
In the Fermi-surface term~\eqref{eq:spin_green2b} that involves both the retarded and the advanced Green's functions,
we have replaced $\hat{v}^{y}(\bm{k})$ and $\hat{s}_{z}$ with $\hat{V}^{y}(\epsilon, \bm{k})$ and $\hat{S}_{z}(\epsilon)$, respectively.
These renormalized vertices, diagrammatically represented in Figs.~\ref{fig:diagram}(b) and \ref{fig:diagram}(c), are obtained by solving
\begin{subequations} \begin{align}
  \hat{V}^{y}(\epsilon, \bm{k})
  = & \hat{v}^{y}(\bm{k}) + n_{\mathrm{i}} v_{\mathrm{i}}^{2}
  \int \frac{d^{2} k^{\prime}}{(2 \pi)^{2}} \notag \\
  & \times \hat{G}^{\mathrm{R}}(\epsilon, \bm{k}^{\prime}) \hat{V}^{y}(\epsilon, \bm{k}^{\prime}) \hat{G}^{\mathrm{A}}(\epsilon, \bm{k}^{\prime}), \label{eq:green5a} \\
  \hat{S}_{z}(\epsilon)
  = & \hat{s}_{z} + n_{\mathrm{i}} v_{\mathrm{i}}^{2}
  \int \frac{d^{2} k^{\prime}}{(2 \pi)^{2}} \notag \\
  & \times \hat{G}^{\mathrm{A}}(\epsilon, \bm{k}^{\prime}) \hat{S}_{z}(\epsilon) \hat{G}^{\mathrm{R}}(\epsilon, \bm{k}^{\prime}). \label{eq:green5b}
\end{align} \label{eq:green5}\end{subequations}
For the bare velocity vertex $\hat{v}^{y}(\bm{k}) = \hbar k_{y}/m + \alpha \sigma_{x}$ and spin vertex $\hat{s}_{z} = (\hbar/2) \sigma_{z}$,
the renormalized vertices are $\hat{V}^{y}(\epsilon, \bm{k}) = \hbar k_{y}/m + \alpha V^{yx}(\epsilon) \sigma_{x}$ and $\hat{S}_{z}(\epsilon) = (\hbar/2) S_{z}^{z}(\epsilon) \sigma_{z}$, respectively.

In the limit of $\Gamma_{0} \rightarrow +0$, Eq.~\eqref{eq:spin_green2a} reproduces the SO magnetic susceptibility~\eqref{eq:rashba3c} obtained by the Bloch formula.
To neglect the vertex corrections, we only have to put $\hat{V}^{yx}(\epsilon) = \hat{S}_{z}^{z}(\epsilon) = 1$,
and Eq.~\eqref{eq:spin_green2b} reproduces Eq.~\eqref{eq:rashba3d} by identifying $\eta \leftrightarrow 2 \Gamma(\epsilon)$.
When we take into account the vertex corrections, we reproduce~\cite{PhysRevB.70.041303,suppl}
\begin{equation}
  V^{yx}(\epsilon)
  =
  \begin{cases}
    0 & (\epsilon > 0) \\
    -2 \epsilon/m \alpha^{2} & (\epsilon < 0)
  \end{cases}, \label{eq:green11}
\end{equation}
and $S_{z}^{z}(\epsilon) = 1$~\cite{Edelstein1990233}.
Then, the correct spin accumulation at $T = 0$ becomes~\cite{suppl}
\begin{align}
  \langle \Delta \hat{s}_{z} \rangle^{(1, 1, \mathrm{I})}(\Omega, \bm{Q})
  = & \frac{-\hbar}{2 \Gamma(\mu)} \times \left(-\frac{q}{8 \pi}\right) (i Q_{x}) E_{y}(\Omega, \bm{Q}) \notag \\
  & \times
  \begin{cases}
    1 & (\mu > 0) \\
    \sqrt{1 + 2 \mu/m \alpha^{2}} & (\mu < 0)
  \end{cases}. \label{eq:spin_green5}
\end{align}
This equation is another main result.
The spin accumulation is nonzero even in the case where the chemical potential is below the Rashba crossing.
We emphasize again that the SH conductivity vanishes
in our setup~\cite{PhysRevB.70.041303,PhysRevLett.93.226602,PhysRevB.71.033311,PhysRevB.71.245318,PhysRevB.73.049901,PhysRevB.71.245327,PhysRevLett.96.056602}.
If we consider the diffusion process, the spin accumulation decays in the scale of the spin-diffusion length~\cite{PhysRevB.98.174422}
as in the experimental~\cite{Kato1910} and theoretical results~\cite{PhysRevB.70.195343,PhysRevLett.95.046601,PhysRevB.72.081301,PhysRevB.73.075303,RASHBA200631}.

\paragraph{Discussion.}
First, let us discuss the directions of the spin and electric field.
The second term of the Bloch formula~\eqref{eq:spin_bloch3d} involves the spin $s_{na}$ and the orbital magnetic moment $m_{nk}$.
Since these two are parallel to each other,
the spin accumulation takes the form of $\langle \Delta \hat{s}_{a} \rangle(0, \bm{Q}) \propto [i \bm{Q} \times \bm{E}(0, \bm{Q})]_{a}$,
more precisely, $\langle \Delta \hat{s}_{a} \rangle(0, \bm{Q}) \propto [i \bm{Q} \times \langle \Delta \hb{J} \rangle(0, \bm{Q})]_{a}$
considering the boundary effect as argued in the Introduction.
Thus, the direction of the spin accumulation is consistent with the typical scenario of the SH effect.

Second, we discuss a relation between our results and the previous results on the SH conductivity.
By multiplying $(-i \Omega)$ to Eq.~\eqref{eq:spin_bloch3a}, we obtain the time derivative of the spin expectation value.
If we take the limits of $\Omega \rightarrow 0$ and $\eta \rightarrow +0$ in the arbitrary order, we may obtain
\begin{align}
  (-i \Omega) & \langle \Delta \hat{s}_{a} \rangle(\Omega, \bm{Q}) \notag \\
  = & \alpha_{a}^{\phantom{a} j} E_{j}(\Omega, \bm{Q})
  - \left[\frac{\hbar}{\eta} \gamma_{a}^{\phantom{a} ij (\mathrm{I})} + (\gamma_{a}^{\phantom{a} ij (\mathrm{II})} + \epsilon^{ijk} \chi_{ak}^{\mathrm{so}})\right] \notag \\
  & \times (i Q_{i}) E_{j}(\Omega, \bm{Q}). \label{eq:discussion1}
\end{align}
Here, we have used Faraday's law, $(-i \Omega) \bm{B}(\Omega, \bm{Q}) = -(i \bm{Q}) \times \bm{E}(\Omega, \bm{Q})$.
Since the second term is the divergence, we can read the spin (Hall) conductivity as
\begin{align}
  \tilde{\sigma}_{sa}^{\phantom{sa} ij}
  = & \frac{\hbar}{\eta} \gamma_{a}^{\phantom{a} ij (\mathrm{I})} + (\gamma_{a}^{\phantom{a} ij (\mathrm{II})} + \epsilon^{ijk} \chi_{ak}^{\mathrm{so}}) \notag \\
  = & \frac{\hbar}{\eta} \gamma_{a}^{\phantom{a} ij (\mathrm{I})}
  - \frac{q}{\hbar}
  \sum_{n} \int \frac{d^{d} k}{(2 \pi)^{d}} \notag \\
  & \times [s_{na}^{\phantom{na} j} \partial_{k_{i}} \epsilon_{n} f^{\prime}(\epsilon_{n}) + \epsilon^{ijk} b_{nak} f(\epsilon_{n})]. \label{eq:discussion2}
\end{align}
This formula is consistent with the SH conductivity of the conserved spin current~\cite{suppl} proposed in Refs.~\cite{PhysRevLett.96.076604,PhysRevB.77.075304}.
Furthermore, the St\v{r}eda formula between the SH conductivity and the SO magnetic susceptibility~\cite{PhysRevLett.97.236805} is obvious in this formalism.
Equation~\eqref{eq:discussion2} can be interpreted that the SH conductivity is not related to the spin accumulation in the case of the nonzero SO magnetic susceptibility.
However, when the vertex corrections are taken into account, Eq.~\eqref{eq:discussion2} no longer holds,
and the SH conductivity is not related to the spin accumulation regardless of the presence or absence of the SO magnetic susceptibility.

Third, we mention first-principle calculations of the spin accumulation, which is obtained by the product of $-\hbar/\eta$ and Eq.~\eqref{eq:spin_bloch3d}.
Although we have treated $\eta \rightarrow +0$ as a constant, it is better to take $\eta$ proportional to the density of states.
This choice is justified for $\delta$-function nonmagnetic disorder within the first Born approximation, although the vertex corrections cannot be taken into account.
To do so, the Korringa-Kohn-Rostoker formalism combined with the coherent potential approximation is useful as demonstrated in the context of the SH conductivity~\cite{PhysRevLett.106.056601}.

In Ref.~\cite{PhysRevB.98.174422}, one of the authors calculated the spin response to the electric-field gradient for the Rashba model
using the first-order perturbation theory with respect to the Rashba SO coupling $\alpha$.
It was found that the response vanishes for the uniform $\alpha$ and is nonzero only when $\alpha$ is nonuniform.
Here, we have calculated the same response nonperturbatively and obtained the nonzero response for the uniform $\alpha$.
In fact, Eq.~\eqref{eq:spin_green5} is universal, i.e., independent of $\alpha$, apart from the imaginary part of the self-energy, which cannot be captured by the perturbation theory.

\paragraph{Summary.}
We have calculated the spin response to the electric-field gradient, which naturally appears at the boundaries.
First, we have derived the Bloch formula~\eqref{eq:spin_bloch4} assuming the phenomenological relaxation time.
We have also calculated the response for the uniform Rashba model with $\delta$-function nonmagnetic disorder
using the first-order Born approximation and corresponding ladder-type vertex corrections.
Although the SH conductivity vanishes~\cite{PhysRevB.70.041303,PhysRevLett.93.226602,PhysRevB.71.033311,PhysRevB.71.245318,PhysRevB.73.049901,PhysRevB.71.245327,PhysRevLett.96.056602},
the spin response~\eqref{eq:spin_green5} is nonzero as observed experimentally~\cite{Kato1910}.
This theory enables us to calculate the spin accumulation in the SH effect without imposing the open boundary conditions or attaching the leads
and, hence, can be implemented in first-principles calculations for real materials.

%--- Back Matter
\begin{acknowledgments}
  \paragraph{Acknowledgments.}
  A.S. thanks J.~Fujimoto for fruitful discussions on the diagram technique.
  We also thank E. I.~Rashba for informing us of Refs.~\cite{RASHBA200631,Bychkov1984}.
  This work was supported by the Japan Society for the Promotion of Science KAKENHI (Grants No.~JP18K13508, No.~JP21H01816, and No.~JP21H01034).
\end{acknowledgments}
\end{document}